\documentstyle[aps]{revtex}

\begin{document}

\author{U\v{g}ur T{\i}rnakl{\i}
\thanks{e-mail: tirnakli@sci.ege.edu.tr}$^,$\thanks
{TUBITAK M\"{u}nir Birsel Foundation Fellow}, 
Fevzi B\"{u}y\"{u}kk{\i}l{\i}\c{c}, Do\v{g}an Demirhan \\
Department of Physics, Faculty of Science, \\
Ege University 35100, Bornova \.{I}zmir-TURKEY}

\title{Some bounds upon nonextensivity parameter
using the approximate generalized distribution functions}

\date{\today}

\maketitle

\begin{abstract}
In this study, approximate generalized quantal distribution functions 
and their applications, which appeared in the literature so far, 
have been summarized. Making use of the generalized Planck radiation law, 
which has been obtained by the authors of the present manuscript 
[Physica A240 (1997) 657], some alternative bounds for nonextensivity
parameter $q$ has been estimated. It has been shown that these results are 
similar to those obtained by Tsallis et al. [Phys.Rev. E52 (1995) 1447] and 
by Plastino et al. [Phys.Lett. A207 (1995) 42].\\

\noindent PACS number(s): 05.20.-y, 05.30.Jp
\end{abstract}

\vspace{1.5cm}

\section{Introduction}
Boltzmann-Gibbs (BG) statistics provides a very powerful and suitable 
tool for handling physical systems where thermodynamic extensivity 
(additivity) holds. Apparently, this condition itself points out that 
BG formalism has some restrictions. These could be stated as follows: 
(i) the spatial ranges of the microscopic interactions must be short, 
(ii) the time range of the microscopic memory must be short and 
(iii) the system must evolve in an Euclidean-like,nonfractal, space-time. 
Whenever one 
and/or the others of these restrictions are violated, BG formalism fails 
to study the system under consideration. This kind of systems are said to 
be nonextensive, therefore a nonextensive formalism of statistics is 
needed for them. In the recent years, an increasing tendency towards 
nonextensive formalisms keeps growing along two lines : 
Quantum-Group-like approaches [1] and Tsallis Thermostatistics (TT) [2], 
to which this paper is dedicated.

TT, which has been proposed by C. Tsallis in 1988, is based upon 
two basic axioms:

(a) the entropy is given by

\begin{equation}
S_q=-k\frac{1-\sum_{i=1}^{W}p_i^q}{1-q}
\end{equation}

\noindent where $k$ is a positive constant, $p_i$ is the probability of the 
system to be in a microstate, $W$ is the total number of configurations and 
$q$ is a real parameter.

(b) the $q$-expectation value of an observable $O$ is given by

\begin{equation}
\left<O\right>_q=\sum_{i=1}^{W}p_i^q O_i
\end{equation}

\noindent It is worthwhile to note that TT contains BG statistics as a 
special case in the $q\rightarrow 1$ limit. Extensive Shannon entropy and 
standard definition of the expectation value are recovered in this limit, 
thus $(1-q)$ could be interpreted as the measure of the lack of extensivity 
of the system.

In fact, it is plausible to classify the works on this subject in the
following manner: (i) Some of the works have been devoted to the
generalization of the conventional concepts of thermostatistics so far, in
order to see whether this nonextensive formalism allows an appropriate
extension of these concepts. (ii) After these efforts, TT has been applied to
some physical systems [3] where BG formalism is known to fail, and its
success on these lines accelerates new attempts nowadays. (iii) Finally, the
last area of interest, which is very active, originates from the fact that
the physical meaning of the $q$-index has only now started to be clarified. 
Efforts on this field accumulate along two main streams; one of which is the 
study of dynamical systems [4] in order to find out the connection between 
the $q$-index and the dynamical parameters of the system, whereas the second 
one is the attempts of establishing some bounds upon $q$ by studying, 
in general, astrophysical problems such as the cosmic microwave background 
radiation, Planck law [5,6,7] and the early Universe [8,9]. In this study, 
our main goal is to contribute to this stream by estimating some alternative 
bounds upon $q$.

Before closing this Section, it could be mentioned that the reader 
can find a detailed review of TT with its important properties 
in [10], moreover, a full bibliography of TT is now 
available in [11].
 
\section{Generalized Quantal Distribution Functions}

Amongst the generalizations of the concepts of thermostatistics in 
the frame of TT, quantal distribution functions have been studied for 
the first time by our group [12,13]. In [12], using the statistical 
weights, the fractal inspired entropies of quantum gases have been 
deduced and the generalized distribution functions have been found by 
introducing the constraints related to the statistical properties of the 
particles. In [13], Tsallis entropy has been used for the same 
purpose and the same results for the generalized distribution functions 
have been reproduced but in this case the constraints have been imposed 
in the calculations of the partition functions.

Very recently, these distribution functions have been reobtained by 
Kaniadakis and Quarati [14], using a completely different technique. Since 
the classical statistical distributions could be viewed as stationary 
states of a linear Focker-Planck equation, analogously, the authors have 
shown that [15] quantal distributions can be obtained, by means of a 
semiclassical approach, as stationary states of a nonlinear Focker-Planck 
equation. They have found that; within this kinetic approach, a family of 
classical and quantal distributions can be derived when the drift and the 
diffusion coefficients are polynomials of $2M+1$ and $2N$ degrees, 
respectively, in the velocity variable. A particular and well-defined 
distribution corresponds to each pair of $(M,N)$, i.e. $(0,1)$ gives the 
generalized distribution functions which are identical with the results 
of [12,13]. It is also worth mentioning that some aspects of generalized 
quantum statistics can be found in [16].

The generalized quantal distribution functions, derived independently in 
these three papers, are given by 

\begin{equation}
\left<n_r\right>_q=\frac{1}{\left[1-(1-q)\beta (\epsilon_r-\mu)\right]^
{1/(q-1)}+\kappa}
\end{equation}

\noindent where $\kappa$ is $+1$ and $-1$ for Fermi-Dirac and 
Bose-Einstein distributions, respectively, $\mu$ being the chemical
potential.  

It is worth emphasizing that these expressions for the quantal 
distribution functions are not exact results due to the approximate 
scheme used in the calculations. This scheme is generically based on 
the factorization of the generalized grand canonical distribution function 
and the concomitant $q$-partition function as if they were extensive 
quantities, due to the fact that a closed, manageable form for them cannot 
be found without making such approximations. The authors of the present 
manuscript have recently discussed the nature of this approximation in [7] 
more clearly and showed that these 
results give upper or lower bounds to the exact results with respect to 
the values of $q$-index. The role of this approximation is very similar 
to the role of the mean-field theory approximation used in the theory of 
phase transitions. This approximate scheme has also been used in another 
problem [17] and it is seen that it gives similar results with the 
prior works on the same subject.  

\section{Applications of the Generalized Quantal Distribution Functions}

Although the generalized distribution functions are first introduced 
in 1993, there has been no attempt to apply them to the physical systems 
until a recent effort performed by Pennini et al. [18]. The authors dealt 
with $N$ independent particles (fermions and bosons), distributed among 
$M$ single particle levels of single particle energies $\epsilon_i$ 
$(i=1,...,M)$. In order to obtain 
the exact result, the authors maximized the Tsallis entropy with suitable 
constraints and found the exact result of an extremely simple model, 
which is the simplest conceivable situation for fermions where $M=2$, 
since otherwise the non-separability of the partition function does not 
allow to find a close, analytical expression for arbitrary $q$ values 
without making an approximation like the one we discussed above. 
The average fermion occupation number of the lowest single particle 
level versus inverse temperature with $q=0.7$ is given in Fig. 1 of 
[18]. Pennini et al. performed the same calculations for bosons as 
well with $M=2$. The average boson occupation number of the lowest single 
particle level with $q=0.7$ is given in Fig. 3 of [18]. In both figures, 
it is clearly seen that our approximate scheme provides bounds to the exact 
results (since Pennini et al. used $q<1$, the bound appears as a lower 
bound). This study also suggests that the quality of our approximation
deteriorates as the number of particles of the system increases, however, 
a very recent effort by Wang and Le M\'{e}haut\'{e} [19] showed clearly 
that there exist a temperature interval where the ignorance of the 
approximation is significant, but outside this interval, the results 
of this approach can be used with confidence. In addition, they verified 
that the magnitude of this interval remains {\it constant} with the 
increase of the number of particles, a rather nice result in the favour 
of the approximate scheme, contrary to the result of Pennini et al.

The second application of the generalized distribution functions has 
recently been performed by the authors of this manuscript [7]. 
Two years ago, the generalization of the Planck law was obtained by 
Tsallis et al. [5] for $q\cong 1$, in order to see whether present cosmic 
background radiation is (slightly) different from the Planck radiation 
law due to the long-range gravitational influence. We have advanced an 
alternative approach to this generalization by using one of the 
generalized distribution functions, namely generalized Planck 
distribution. Once again, as expected, our approximate scheme 
shows itself as an upper or lower bound to the exact result, depending 
on the value of $q$. This fact is verified clearly in the figure of 
[7] by plotting the blackbody photon energy density per unit 
volume versus $h\nu /kT$ in the frame of [7] and [5]. 
We have also been able to obtain all the generalized laws, which was 
found by Tsallis et al., (i.e., generalized Planck, Rayleigh-Jeans, 
Stefan-Boltzmann and Wien laws) and noticed that our approach is 
simpler than that of Tsallis et al. since the procedure we have used 
is completely the same as that followed in any standard textbook 
of Statistical Physics for the derivation of the standard Planck law and 
also our approximation which leads bounds seems to be more general, 
since not necessarily $q\cong 1$.

\section{Estimation of Some Bounds Upon $q$}

In [7], the generalized Planck radiation law is given by 

\begin{equation}
D_q(\nu)\simeq\frac{8\pi k^3T^3}{c^3h^2}\frac{x^3}{\left[1-(1-q)x\right]^
{\frac{1}{q-1}}-1}
\end{equation}

\noindent where $x\equiv h\nu /kT$. For our purpose, it is 
straightforward to expand this expression to first order in $(1-q)$, 
which reads

\begin{equation}
D_q(\nu)\simeq\frac{8\pi k^3T^3}{c^3h^2}\left[\frac{x^3}{e^x-1}-
\frac{1}{2}(1-q)\frac{x^5e^x}{\left(e^x-1\right)^2}\right]\;\; .
\end{equation}

\noindent For a comparison, eq(5), together with the results of [7] and
[5] for $D_q(\nu)$, have been illustrated in the Figure.

The value of the frequency $\nu_m$, which makes $D_q(\nu)$ maximum, can
easily be found by maximizing eq.(5):

\begin{equation}
x_m\equiv h\nu_m/kT\simeq a_1 + a_2 (q-1)
\end{equation}

\noindent where $a_1\simeq 2.821$ and $a_2\simeq 4.928$. Substitution of
this result in eq.(5) immediately yields

\begin{equation}
D_q(\nu_m)\frac{c^3h^2}{8\pi k^3T^3}\simeq d_1 + d_2 (q-1)
\end{equation}

\noindent with $d_1\simeq 1.421$ and $d_2\simeq 10.677$. Lastly, we verify
that, for $\nu\simeq \nu_m$,

\begin{equation}
\frac{D_q(\nu)}{D_q(\nu_m)}\simeq 1- B \left(\frac{\nu}{\nu_m}-1\right)^2
\end{equation}

\noindent where 

\begin{equation}
B\simeq b_1 - b_2 (q-1)
\end{equation}

\noindent with $b_1\simeq 1.232$ and $b_2\simeq 2.322$. From these results,
the following expression can easily be written:

\begin{equation}
q-1\simeq \frac{b_1 - B^{expt}}{b_2}
\end{equation}

\noindent which is nothing but an alternative estimation of a bound upon $q$,
similar to that of given in [5] by Tsallis et al. (here, $B^{expt}$ stands
for the experimental value of $B$). Note that although, at first sight, 
eq.(10) seems to be the same with eq.(24) of [5], in fact they are
different since the coefficient $b_2$ is not equal to the one given in [5].
Strickly speaking, it must be pointed out that all of the coefficients which
belong to the nonextensivity part (namely, $a_2$, $b_2$, $d_2$) differ from
those given in [5].

In the remainder of this manuscript, we focus our attention to another bound
upon $q$, which was established by Plastino et al. [6] with the help of the
expression of $D_q(\nu)$ given in [5] and the experimental value of the
Stefan-Boltzmann constant. Following the same procedure in [6], but using
eq.(5) for $D_q(\nu)$, we establish another bound upon $q$.

Let us write down the total emitted power per unit surface,
\begin{equation}
P_q=\int_0^{\infty}D_q(\nu) d\nu = \sigma _q T^4
\end{equation}

\noindent where $\sigma _q$ is the generalized Stefan-Boltzmann constant,
which is given by 
\begin{equation}
\sigma _q\cong \frac{8\pi k^4}{c^3h^3} \int_0^{\infty}
\left[\frac{x^3}{e^x-1}-\frac{1}{2}(1-q)\frac{x^5e^x}{\left(e^x-1\right)^2}
\right] dx\;\; .
\end{equation}

\noindent If we denote the integral appearing in the above expression by $I$, 
the explicit numerical solution reads

\begin{equation}
I=6.4939 - 62.2157 (1-q)
\end{equation}

\noindent therefore,

\begin{equation}
\sigma _q=5.67\times 10^{-8} - 3.49\times 10^{-7} (1-q)\;\; .
\end{equation}

\noindent Using this, together with

\begin{equation}
\sigma _{Planck}=5.67051(19)\times 10^{-8}\;\; W \;m^{-2} K^{-2}
\end{equation}

\noindent and the '3 sigma' rule [6]

\begin{equation}
\left|\frac{\left[\left(\delta\sigma_{Planck}\right)^2 + 
\left(\delta\sigma_q\right)^2\right]^{1/2}}{\sigma_q}\right|
\leq 4.1\times 10^{-4}\;\;\; ,
\end{equation}

\noindent it is possible to find a bound upon $q$:

\begin{equation}
\left|q-1\right|\leq 0.41\times 10^{-4}\;\;\; ,
\end{equation}

\noindent which is similar to that obtained by Plastino et al. [6].

\section{Conclusions}
In this study, we have concentrated on estimating new bounds upon
nonextensivity parameter $q$. For this purpose, the generalized Planck 
radiation law has been used and following the same procedure of [5] and [6], 
two new bounds for $q$ has been estimated. These results are, as 
expected, consistent with the previous ones, therefore the same conclusions 
are valid for the present case.

On the other hand, very recently we have realized that the present effort 
might motivate some other works on the early Universe. More precisely, 
Torres et al. [8], within an early Universe scenario, calculated the 
deviation in the primordial Helium abundance and obtained a bound for $q$. 
Moreover, Torres and Vucetich studied the primordial neutron to baryon ratio 
in a cosmological expanding background [9] and within this context, 
they found another bound upon $q$. In both works, the generalized 
mean value of the particle number operator $\left<n\right>_q$ has been used. 
Hence, it is clear that these works could be handled again by using the 
suitable definition of $\left<n\right>_q$ given in eq.(3).

Summing up, we applied one of the generalized quantal distribution functions
to the microwave background radiation in order to estimate new bounds upon $q$
and further attempts on this line would highly be welcomed. 

\section{Acknowledgments}

We are indebted to A.R. Plastino for valuable remarks on ref.[6] and to 
D.F. Torres for sending refs.[8,9] prior to publication.
We would like to thank Ege University Research Fund for their partial
financial support under the Project Number 97 FEN 025.


\newpage

\section*{Figure Caption}

\vspace{2cm}

\noindent {\bf Figure :} Blackbody photon energy density per unit volume 
versus $h\nu/kT$ in the frame of ref.[5], ref.[7] and this work, 
for $q=0.95$, $q=1$ and $q=1.05$.

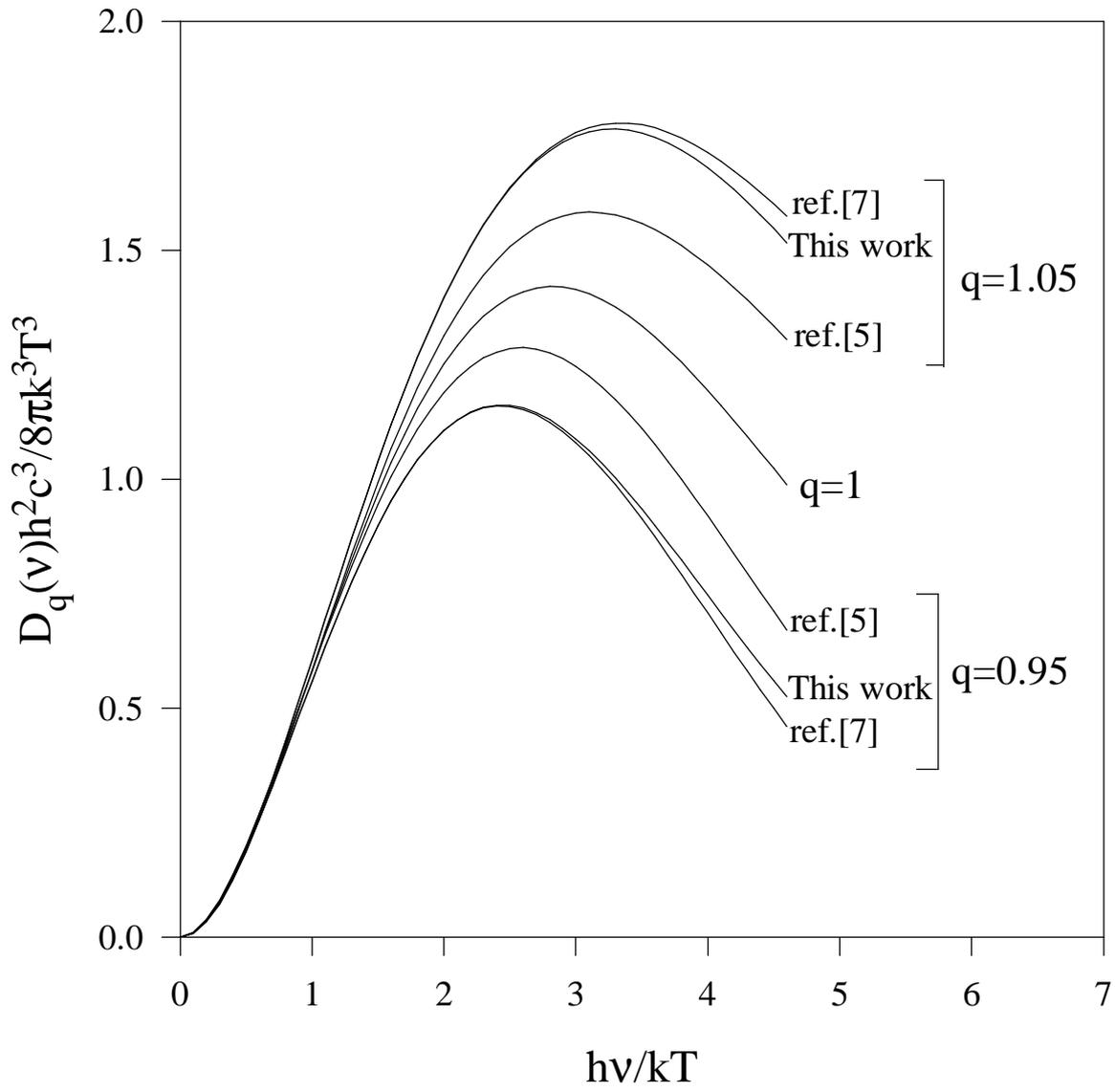
\end{document}